\documentclass[twocolumn,aps,prx,superscriptaddress]{revtex4-2}

\usepackage{chngcntr}

\usepackage{graphicx}
\usepackage{color}
\usepackage{amsmath}
\usepackage{enumitem}
\usepackage{amssymb}
\usepackage{hyperref}
\usepackage{cancel}
\usepackage{ulem}
\usepackage{multirow}

\newcommand{\be}{\begin{equation}}
	\newcommand{\ee}{\end{equation}}

\newcommand{\bea}{\begin{eqnarray}}
	\newcommand{\eea}{\end{eqnarray}}

\newcommand{\lb}{\left[}
\newcommand{\rb}{\right]}
\newcommand{\lp}{\left(}
\newcommand{\rp}{\right)}

\renewcommand{\vec}[1]{{\boldsymbol #1}}

\begin{document}
	\title{Signatures of Cooper pair dynamics and quantum-critical superconductivity in tunable carrier bands}
	\author{Zhiyu Dong, Patrick A. Lee, Leonid S. Levitov}
	\affiliation{Department of Physics, Massachusetts Institute of Technology, Cambridge, MA 02139}

\begin{abstract}
Different superconducting pairing mechanisms are markedly distinct in the underlying Cooper pair kinematics. Pairing interactions mediated by quantum-critical soft modes are dominated by highly collinear processes, falling into two classes: forward scattering and backscattering. In contrast, phonon mechanisms have a generic non-collinear character. We show that the type of kinematics can be identified by examining the evolution of superconductivity when tuning the Fermi surface geometry. We illustrate our approach using recently measured phase diagrams of various graphene systems. Our analysis unambiguously connects the emergence of superconductivity at ``ghost crossings'' of Fermi surfaces in distinct valleys to the pair kinematics of a backscattering type. Together with the observed non-monotonic behavior of superconductivity near its onset (sharp rise followed by a drop), it provides strong support for a particular quantum-critical superconductivity scenario. These findings conclusively settle the long-standing debate on the origin of superconductivity in this system and demonstrate the essential role of quantum-critical modes in superconducting pairing. Moreover, our work highlights the potential of tuning bands via ghost crossings as a promising means of boosting superconductivity.
\end{abstract}

\date{\today}
%\dates{This manuscript was compiled on \today}
%\doi{\url{www.pnas.org/cgi/doi/10.1073/pnas.XXXXXXXXXX}}
%
%\maketitle
%\thispagestyle{firststyle}
%\ifthenelse{\boolean{shortarticle}}{\ifthenelse{\boolean{singlecolumn}}{\abscontentformatted}{\abscontent}}{}

% If your first paragraph (i.e. with the \dropcap) contains a list environment (quote, quotation, theorem, definition, enumerate, itemize...), the line after the list may have some extra indentation. If this is the case, add \parshape=0 to the end of the list environment.

\maketitle

Superconducting phases occurring in various strongly interacting systems 
\cite{cao2018SC,cao2018insulator,zhou2021superconductivity,zhou2022isospin,zhang2022promotion,kamihara2008iron} are often interpreted by theoretical frameworks that involve quantum-critical pairing \cite{berk1966effect,doniach1966low,layzer1971spin,fay1980coexistence,fernandes2014drives,Mazin2008unconventional,Mazin2009pairing,
lu2022correlated,dong2022spin,dong2021superconductivity,chatterjee2022inter,
you2022kohn}. Yet, delineating these experimentally from the more conventional scenarios has not always been easy. %In this vein, superconductivity recently 
Superconductivity (SC) observed in moir\'e and non-moir\'e graphene at the onset of electronic orders, where soft spin and valley collective modes can mediate  pairing\cite{dong2022spin,chatterjee2022inter,lu2022correlated,you2022kohn,
dong2021superconductivity,wang2021topological}, is an appealing setting for understanding the telltale signatures of different pairing mechanisms. Pairing with nonzero angular momentum can often be identified from the dependence on the applied magnetic field. In this vein, are there easily identifiable signatures of superconductivity driven by quantum-critical soft modes? 

Tuning the band parameters in correlated electron systems through the quantum-critical point (QCP) in order to gain insight into the nature of superconductivity has been a subject of wide interest. 
%\addQ{Yet, }it is experimentally very challenging to adjust the band structure by anything more than relatively subtle perturbations.
In most cases, modifying the band structure beyond subtle perturbations is extremely difficult to achieve experimentally.
Nevertheless, the dependence on an applied strain has been used to reveal %studied in systems such as  Sr$_2$RuO$_4$ and iron-based superconductors. It reveals 
the impact of the van Hove points on the superconductivity in Sr$_2$RuO$_4$\cite{hicks2014strong,steppke2017strong,barber2018resistivity,nomura2002roles,bergemann2003quasi}, and the competition between nematic order and superconductivity in iron-based superconductors\cite{chu2010plane,chu2012divergent,kuo2016ubiquitous}. In the $\kappa$-phase organic superconductors\cite{elsinger2000kappa} and heavy fermion systems such as CeCoIn$_5$\cite{sidorov2002superconductivity} and UPt$_3$\cite{hayden1992antiferromagnetic,sauls1994order}, the role of interaction and %the  relation between interactions, 
correlations %and superconductivity 
is probed by pressure dependence of the superconductivity. These findings have triggered considerable theoretical interest \cite{paul2017lattice,fernandes2010effects,mckenzie1998strongly,
monthoux1999p,monthoux2001magnetically,joynt2002superconducting}.
%\addQ{These measurement lead to theoretical study of signature of superconductivity in tunable systems\cite{paul2017lattice,fernandes2010effects,mckenzie1998strongly,monthoux1999p,monthoux2001magnetically,joynt2002superconducting}.}  
%To the contrary, 

Unlike previously studied systems, in graphene-based superconductors the Fermi surfaces are widely tunable\cite{cao2018SC,cao2018insulator,zhou2021superconductivity,zhou2022isospin,zhang2022promotion}. This tunability, as we will see, opens new avenues for probing the nature of pairing through linking it to %the general character of 
the Cooper pair scattering kinematics. The latter are known to be highly collinear for superconductivity (SC) assisted by incipient electronic orders and driven by soft quantum-critical modes\cite{
	lu2022correlated,chatterjee2022inter,
	you2022kohn}.
%	\addQ{Specifically, the Cooper pair scattering kinematics in exotic superconductivity scenarios is typically of a highly collinear caracter. 
Depending on the mechanism type it falls into two main classes: %can be of one of the %two main typesdisplay strong anisotropy. This anisotropy allows us to categorize the scattering kinematics into 
% two main types: 
	collinear backscattering and forward scattering. 
	The method we introduce below can differentiate between kinematic types by identifying unique features in the evolution of superconducting phases upon adjusting the Fermi surface geometry.
%The method we introduce below can readily distinguish between these kinematic types by linking them to distinctive features in the evolution of  superconducting phases under tuning the Fermi surface geometry.  
%By comparing the results with 
%measured superconducting

\begin{figure}
\centering
\includegraphics[width=0.45\textwidth]{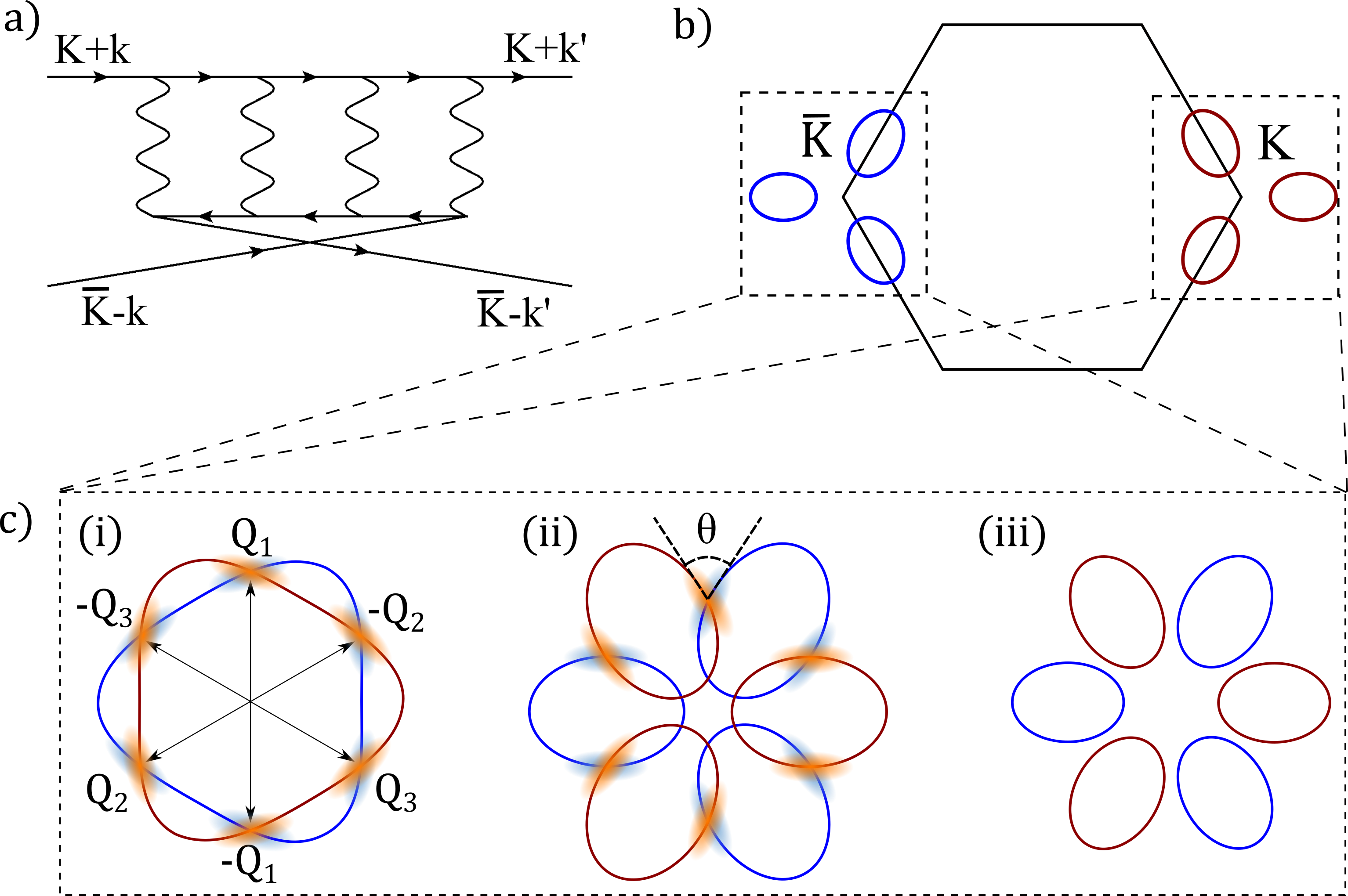}
\caption{ 
	a) A diagrammatic representation of 
	%\sout{backscattering processes mediated by a soft IVC mode ---the quantum-critical pairing interaction that}
	the pairing interaction mediated by isospin mode, %\eqref{eq:Gamma main text}, 
	a soft mode associated with the intervalley phase coherence (IVC). 
	Identifying valleys $\vec K$ and $\bar {\vec K}$ with spin up and down maps this interaction to the paramagnon pairing mechanism mediated by ferromagnetic spin fluctuations\cite{berk1966effect,doniach1966low,layzer1971spin,fay1980coexistence}. 
	At an isospin quantum criticality, this interaction 
	peaks at $\omega+\omega'=0$ and $\vec k+\vec k'=0$ (see \eqref{eq:Gamma main text}), resulting in a backscattering-type Cooper pair dynamics. 
	b) Fermi sea pockets located near valleys $\vec K$ and $\bar {\vec K}$ in BBG bandstructure, which host fermions forming $\vec K$-$\bar {\vec K}$ Cooper pairs. c) Because of backscattering, pairing develops near ``ghost'' crossings of Fermi surfaces $\pm Q_i$ ($i=1,2,3$) found by superimposing the pockets in valleys $\vec K$ (red contours) and $\bar {\vec K}$ (blue contours) by a $\vec K\to \bar {\vec K}$ translation. 
	A single Fermi sea gives non-removable crossings (i), 
	whereas multi-pocket Fermi seas give removable 
	crossings (ii)-(iii), where $\theta$ denotes the angle at the crossing. Transitions between the 
	intersecting and nonintersecting 
	Fermi surfaces (ii) and (iii) induced by tuning the bandstructure terminate the superconducting phases. 
} 
\label{fig:FS and valley-crossing points}
\vspace{-4mm}
\end{figure}

Here, through a detailed quantitative comparison to experimental data obtained by tuning SC in several graphene systems, we 
%LL unambiguously prove
demonstrate 
the occurrence of the collinear backscattering kinematics. Specifically, we find direct evidence linking the onset of superconductivity and the abrupt appearance of ``ghost valley crossings'' between Fermi surfaces in different valleys. This is distinct from conventional ways to stimulate superconductivity %which is tuned by by varying carrier doping or 
by tuning the Fermi level through van Hove points.
%\sout{Conventionally, superconductivity is tuned by varying carrier doping or tuning the Fermi level through van Hove points. Here it is done differently, by tuning through  ``ghost valley crossings''.}
% in these systems. %Meanwhile, as the Cooper pair scattering mediated by different type of soft modes exhibit distinct kinematics, 
	Identification of an abrupt onset of SC with such crossings limits the possible soft modes that can serve as pairing glue, excluding many of the previously considered scenarios and %method 
	pinpointing %the pairing mechanism
	SC driven by the isospin  inter-valley-coherent (IVC) 
	mode pictured in Fig.\ref{fig:FS and valley-crossing points} and discussed below as the most likely %\sout{the only possible type of}
	mechanism. %\addQ{\bf [to strengthen the claim, while making it bulletproof]} % mode that takes on the role of pairing glue in these systems. %, thereby solving a central problem in the field of exotic superconductivity.
	Further evidence for this scenario is provided by a significant enhancement of superconducting $T_{\rm c}$ and a characteristic nonmonotonic behavior at SC onset near ghost valley crossing (see \eqref{eq:Tc main text} and accompanying discussion), which is in good agreement with experimental observations (see Fig.\ref{fig:BBG+WSe2_all}). 

A salient feature of graphene superconductivity that will be important for our analysis is that the two electrons forming a Cooper pair are located in valleys $K$ and $\bar K$ which are related by time reversal symmetry. %Therefore, the backscattering 
Accordingly, Cooper pair kinematics %refers to 
involves valley-conserving scattering of pair states $({\vec K}+{\vec k},\bar {\vec K}-{\vec k}) \to ({\vec K}+\vec k',\bar {\vec K}-\vec k')$ with $\vec k,\vec k'\ll K$. 
%\addQ{with $\vec k'\approx -\vec k$ and $\vec k,\vec k'\ll K$.}  
Because of this property, the only %\sout{known}
type of pairing mechanism that can %\sout{naturally}
generate collinear backscattering kinematics with $\vec k'\approx -\vec k$ is the pairing mediated by isospin fluctuations that are softened and activated at quantum criticality \cite{you2022kohn,chatterjee2022inter,lu2022correlated,dong2021superconductivity,wang2021topological}. Here, isospin refers to spin and valley. This isospin mode arises from the fluctuations of valley order $\langle\psi_{\vec K}^\dagger\psi_{\bar{\vec K}}\rangle$, the quantity describing the inter-valley coherence (IVC)\cite{po2018origin}. 
%Therefore this pairing scenario is also known as ``IVC pairing''. 
To clarify the backscattering nature of the IVC pairing mechanism, we write down the pairing interaction % generated by this scenario diagrammatically as 
shown diagrammatically in Fig.\ref{fig:FS and valley-crossing points}. This is directly analogous to the paramagnon pairing mechanism near a ferromagnetic quantum critical point \cite{berk1966effect,doniach1966low,layzer1971spin,fay1980coexistence}. %involves electron pairs residing in valleys $\vec K$ and $\bar {\vec K}=-\vec K$. 
%	We illustrate this by pairing assisted by 
%	%\sout{the inter-valley-coherent (IVC) order }
%	\addQ{an isospin soft mode activated at the quantum criticality} in two-valley systems such as graphene\cite{you2022kohn,chatterjee2022inter,lu2022correlated,dong2021superconductivity,wang2021topological}. \addQ{Here, isospin refers to spin and valley. This ispspin mode arises from the fluctuation of valley order $\langle\psi_{\vec K}^\dagger\psi_{\bar{\vec K}}\rangle$ which is known as inter-valley coherence(IVC)\cite{po2018origin}. Therefore, below we refer to the pairing mediated by this soft mode as ``IVC pairing".} 
%This pairing interaction, which is directly analogous to that near a ferromagnetic quantum critical point \cite{fay1980coexistence}, involves electron pairs residing in valleys $\vec K$ and $\bar {\vec K}=-\vec K$. The IVC pairing vertex function, shown in Fig.\ref{fig:FS and valley-crossing points}, describes scattering of electron pairs between states $({\vec K}+{\vec k},\bar {\vec K}-{\vec k})$ and $({\vec K}+\vec k',\bar {\vec K}-\vec k')$. 
Standard analysis [see \cite{chatterjee2022inter,lu2022correlated,you2022kohn} and
\cite{SM}, Sec.\ref{sec:pairing interaction}] 
%\addQ{[see also \cite{chatterjee2022inter,lu2022correlated,you2022kohn}]} 
yields
\be\label{eq:Gamma main text}
\Gamma_{\omega \vec k, \omega' \vec k'} 
=  \frac{U}{ \kappa |\omega+\omega'| + l_0^2 (\vec k+ \vec k')^2 + \delta^2 },   
\ee
%that peaks at $\vec k'=-\vec k$ and $\omega'=-\omega$.
where $U$, $\kappa$, $l_0$ are model-specific parameters and $\delta$ denotes the distance to the QCP \cite{SM}. 
%The quantity in \eqref{eq:Gamma main text} is similar in form to the interaction mediated by paramagnon fluctuations occurring at ferromagnetic QCP\cite{fay1980coexistence}. 
% It is important to note that 
Crucially, the  two electrons in a Cooper pair are predominantly scattered from the initial momenta of $(\vec K+\vec k,\bar {\vec K}-\vec k)$ to the final momenta of $(\vec K+\vec k',\bar {\vec K}-\vec k')$ where $\vec k'\approx -\vec k$, namely, backscattering dominates. 
Indeed, the soft mode describing the IVC instability, which mediates pairing, is the particle-hole ladder shown in Fig.\ref{fig:FS and valley-crossing points} for which the momentum transfer is $(\vec K+\vec k)-(\bar {\vec K}-\vec k')$. 
%This can be understood  from Fig.\ref{fig:FS and valley-crossing points}a by noting that the IVC instability is captured by the particle-hole ladder which is characterized by the momentum difference $(\vec K+\vec k)-(\bar {\vec K}-\vec k')$ and is most singular near the IVC ordering vector $\vec K-\bar {\vec K}$. 
Expanding about the ordering vector $2\vec K$ yields a singularity at small $(\vec k+ \vec k')^2$ in \eqref{eq:Gamma main text}. This behavior is distinct from the %is in strong contrast with a second type of 
QCP scenarios where pairing mainly benefits from forward-scattering processes, wherein electrons are scattered by a small angle on the Fermi surface, as, e.g.,  %One well studied example is 
the pairing mediated by nematic fluctuations in iron-based superconductors\cite{klein2020normal,oganesyan2001quantum,
lederer2015enhancement} or %. Another example is 
pairing through interaction renormalized by valley-polarization fluctuations in graphene bilayer\cite{dong2022spin}. In these cases the pairing interaction can be modeled by an expression similar to that in \eqref{eq:Gamma main text} with frequencies and momenta entering as $\omega-\omega'$ and $\vec k-\vec k'$. In this case the interaction peaks at $\vec k'\approx \vec k$ and $\omega'\approx\omega$. 
Therefore, establishing the backscattering pair kinematics %directly proves
strongly supports the IVC pairing mechanism. Since the Fermi surface ghost crossing signature arises generally in the presence of multiple Fermi pockets and tunable bands, this method can be tested in many superconducting systems %\sout{with these properties,}
 such as those found in  transition metal dichalcogenides and graphene multilayers\cite{zhou2021half,zhou2021superconductivity,zhou2022isospin,
		de2022cascade,seiler2021quantum,zhang2022spin,zhang2022promotion}.%\addQ{[\bf This is to replace the text below.]}
		
		Parenthetically, other scenarios may be considered, such as pairing mediated by antiferromagnetic (AFM) fluctuations, where electrons are predominantly scattered between different parts of the Fermi surface by a large AFM ordering momentum. This mechanism is actively studied in iron pnictides\cite{Mazin2008unconventional,Mazin2009pairing}, yet it does not appear relevant for graphene. 

%\addQ{\bf [We have already briefly describe results on page 1, and detailed the description on page 2, so I think we don't need to re-ask the question here.]}
%\sout{Can this difference in the pair scattering kinematics be linked to an observable signature allowing to pinpoint the pairing mechanism and distinguish between different QCP scenarios? In this paper, we address this question by considering the QCP pairing involving mostly backscattering. The signature in question arises in the presence of multiple Fermi pockets and can be tested in tunable bands, such as those found in transition metal dichalcogenides and graphene multilayers}\cite{zhou2021half,zhou2021superconductivity,zhou2022isospin,de2022cascade,seiler2021quantum,zhang2022spin,zhang2022promotion}.

\begin{figure}[t]
\centering
\includegraphics[width=0.45\textwidth]{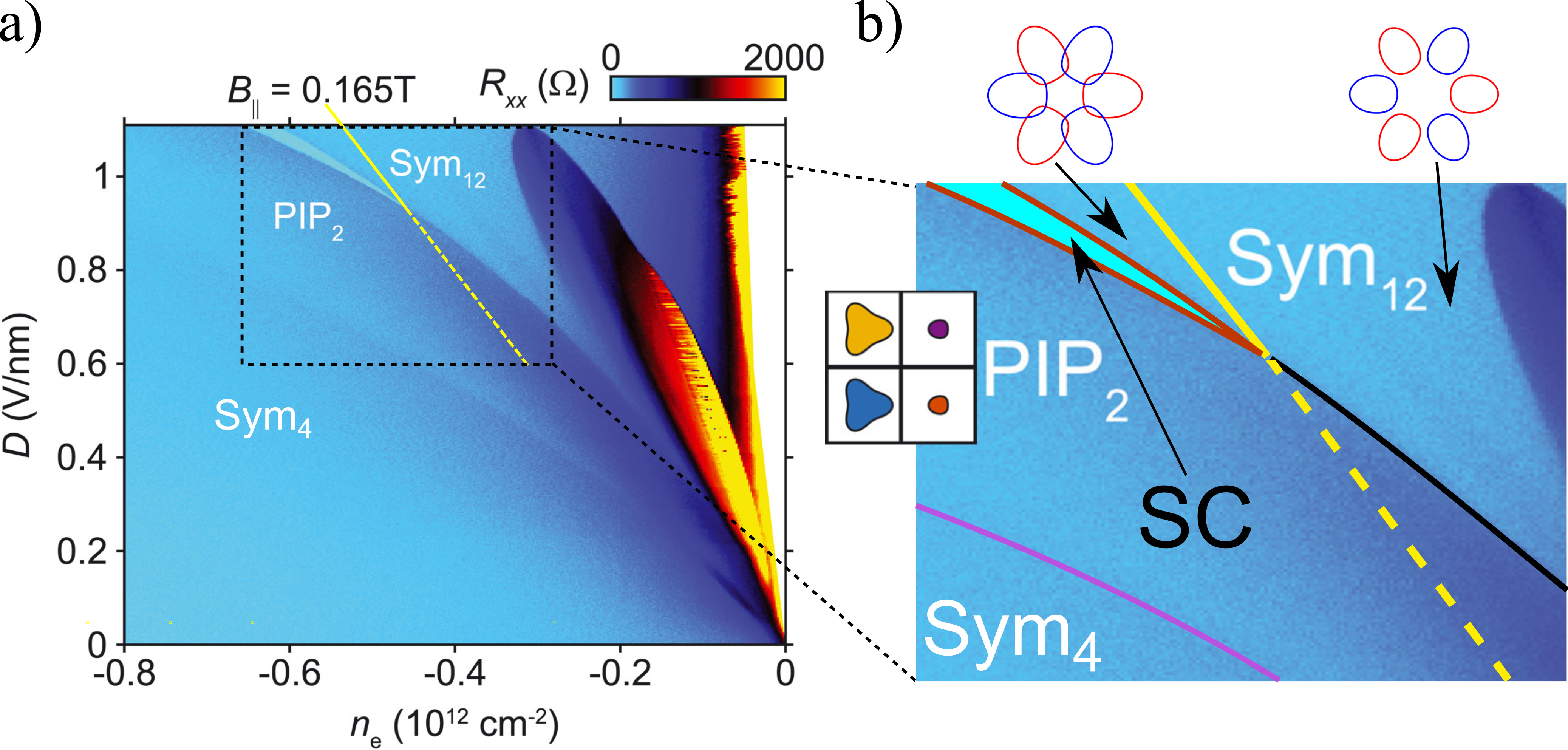}
\caption{
	a) Phase diagram of BBG measured at a finite in-plane magnetic field $B_\parallel= 0.165$T, adapted from Ref.\cite{zhou2022isospin}. 
	The SC phase occurs along the phase boundary between the partially isospin-polarized phase (PIP$_2$) and an isospin-unpolarized phase (Sym$_{12}$). b) 
	A zoom-in of the region near SC %termination point 
	onset in a). Red and black curves mark the phase boundaries.  The solid yellow line, %which is 
	obtained from %band structure calculation 
	the free-particle bands in phase Sym$_{12}$, marks the transition at which the ``ghost'' $\vec K-\bar {\vec K}$ Fermi surface crossings abruptly disappear
	%		between two areas in Sym$_{12}$ phase with distinct valley-crossing behavior. On its left, valley-crossing occurs; on its right, valley-crossing does not occur 
	(see insets in the top row). The dashed yellow line, drawn in the region where the free-particle description does not apply, is a guide to the eye. The measured %termination point 
	emergence point of SC coincides with the %dis
	appearance of the Fermi-surface crossings. The inset on the left, adapted from Ref.\cite{zhou2022isospin}, shows the Fermi surfaces for four isospins in PIP$_2$ phase. %{\bf [???]} 
}
\label{fig:BBG_all}
\vspace{-4mm}
\end{figure}

We will demonstrate the fundamental idea using the setting of Bernal bilayer graphene (BBG) biased by a transverse electric field, a strongly interacting system with a tunable band hosting a superconducting phase\cite{zhou2022isospin}.  A key experimental finding that points to QCP physics is that the SC phase is a sliver that tracks the phase boundary between a  partially-isospin-polarized phase and an unpolarized phase, labeled PIP$_2$ and Sym$_{12}$ in Fig.\ref{fig:BBG_all} following Ref.\cite{zhou2022isospin}. The two QCP scenario types introduced above, involving forward scattering and backscattering, are both viable candidates for this system. The former involves valley-polarization order due to Stoner valley imbalance instability in BBG\cite{dong2022spin}, whereas the latter involves IVC order. The IVC scenario has been considered in RTG  \cite{chatterjee2022inter} and it is straightforward to generalize to BBG as will be shown below. 
%\cite{SM} \addQ{[take out this reference?]}. 
Experiments also indicate a peculiar $B$ dependence of SC which persists in a high in-plane magnetic field $B_\parallel$ and is activated only above a threshold $B_\parallel$. Yet these observations cannot directly distinguish the two QCP scenarios. 

However, there is one observation that so far has escaped attention: The SC sliver only exists on a segment of the PIP$_2$-Sym$_{12}$ phase boundary -- it %terminates 
emerges abruptly %at some point 
upon %reducing 
increasing carrier density along this boundary. The same behavior is found recently in the SC$_1$ phase of BBG/WSe$_2$ but not in RTG. %What makes this behavior particularly intriguing is that the Stoner-instability models that successfully explain the PIP$_2$-Sym$_{12}$ phase boundary do not predict any dramatic changes in e-e interactions along the boundary.
As we shall show, this behavior %makes us 
favors a pairing mechanism that involves backscattering, as opposed to the Stoner instability type proposed earlier which involves forward scattering\cite{dong2022spin}.

%\addLL{We now focus on %show that this termination behavior 
%a link between SC emergence and changes in the Fermi surface geometry, %the appearance of is 
%a signature that distinguishes the backscattering %mechanism 
%pair dynamics from other scenarios.} 
Next, we consider %discuss the analysis of 
the backscattering mechanism for  superconductivity and its relation to %, specifically, the significance of 
ghost crossings. 
As we will see, the pairing gap predominantly opens near the crossings of Fermi surfaces in valleys $\vec K$ and $\bar {\vec K}$ superimposed by a $\vec K\to\bar {\vec K}$ translation. These points, below referred to as ``ghost'' crossings, are illustrated
in Fig.\ref{fig:FS and valley-crossing points} c), where they are labeled $\pm Q_i$ ($i=1,2,3$).
This result directly follows from the back-scattering nature of the pairing interaction \eqref{eq:Gamma main text}, which requires that 
both momenta $\vec k$ and $-\vec k$ are found near the Fermi surface in the same valley. 

\begin{figure}
\centering
\includegraphics[width=0.45\textwidth]{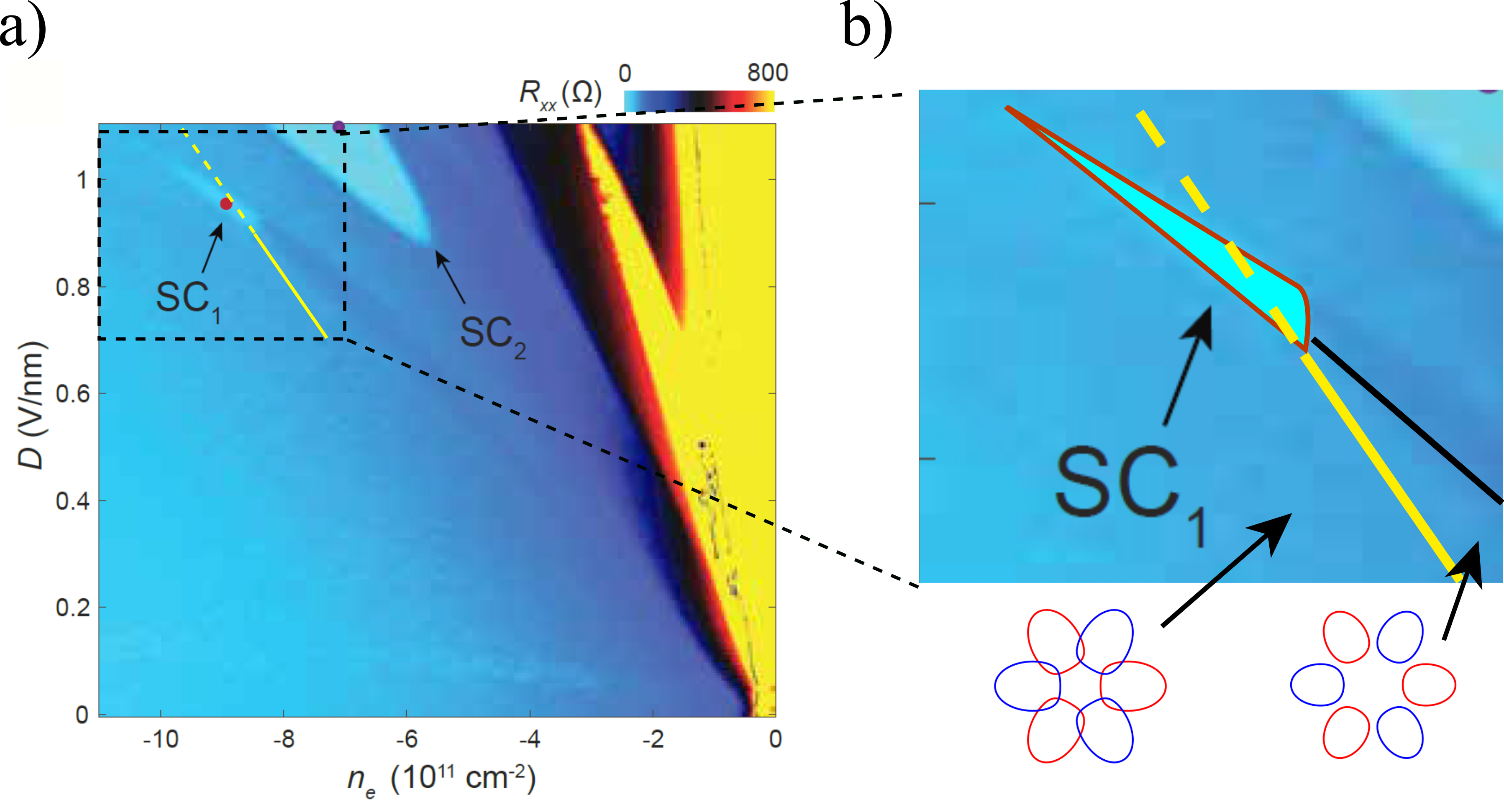}
\caption{a) Phase diagram of WSe$_2$-supported BBG, adapted from Ref.\cite{holleis2023ising}. Here, distinct from BBG, the part below the red and black phase boundaries is the ``vanilla'' phase where no symmetry is broken. The phase above the black phase boundary and SC phase is an ordered phase conjectured in Ref.\cite{holleis2023ising} to be a nematic phase. b) The agreement between theoretically predicted onset of $\vec K$-$\bar {\vec K}$ Fermi surface crossings (solid yellow line) and the measured superconductivity %termination 
emergence point for WSe$_2$-supported BBG. The dashed yellow line, as in Fig.\ref{fig:BBG_all}, is a guide to the eye. 
%The meaning of all notations are the same as in Fig.\ref{fig:BBG_all}. 
	The insets in b) illustrate the overlapping and non-overlapping Fermi surfaces. % under two cases: overlapping (left) and non-overlapping (right). 
	The red and blue curves represent Fermi surfaces in minority isospin species $\vec K\downarrow$ and $\bar{\vec K} \uparrow$, respectively.
	%\addLL{[What about majority species?]}
}
\label{fig:BBG+WSe2_all}
\vspace{-4mm}
\end{figure}

Crucially, these crossings can be switched on and off by varying transverse electric field, an experimental knob tuning the BBG band structure. We anticipate that this change in the bandstructure, illustrated schematically in Fig.\ref{fig:FS and valley-crossing points} c) (ii) and c) (iii), leads to an abrupt emergence %termination 
of SC phase, 
a notable feature observed in BBG (see Fig.\ref{fig:BBG_all}) and WSe$_2$-supported BBG\cite{holleis2023ising} (see Fig.\ref{fig:BBG+WSe2_all}).
This leads to a conjecture that the superconductivity in both systems is dominated by a backscattering pairing mechanism. 
Below we present microscopic analysis for IVC QCP that allows us to verify this conjecture quantitatively by a direct band structure calculation. 
%We note that 
Though for the WSe$_2$-supported BBG the IVC phase is not believed to be stabilized\cite{holleis2023ising,xie2023flavor,wang2023electrical}, %. Nevertheless 
we assume that it may be a competing phase so that IVC fluctuations co-exist with nematic fluctuations, with the IVC pairing channel enhanced by nesting at the ``ghost'' Fermi surface crossing (see below).
%QUESTION:  Do we expect IVC to have higher Tc than nematic because we do not rely on retardation??? } 
Our approach reproduces the measured SC %termination 
emergence points with high accuracy providing strong evidence for pair backscattering. %this pairing mechanism. 
Further, since this behavior cannot be explained by other existing scenarios, such as \cite{dong2022spin,chou2021acoustic,chou2022acoustic,ghazaryan2021unconventional,cea2022superconductivity,cea2023superconductivity,jimeno2022superconductivity}, %we are led to conclude that  
the IVC pairing scenario stands out as the most probable occurrence in a realistic system.
%, most likely, it is the IVC pairing scenario that occurs in a realistic system.

It is also interesting to mention that in the rhombohedral trilayer graphene 
where the Fermi sea is an annulus with both its inner and outer Fermi surfaces looking like the one in Fig.\ref{fig:FS and valley-crossing points} c) (i)\cite{zhou2021half,zhou2021superconductivity}. In this case the ghost crossings remain robust under variation of the electric field and, therefore, we do not expect abrupt emergence or termination of SC phase similar to that seen in BBG. This conclusion is in agreement with the observed phase diagrams \cite{zhou2021superconductivity}. 

Next, we present the essential points of the microscopic analysis.
% of the backscattering mechanism for  superconductivity and, specifically, the significance of ghost crossings. 
Due to the observation that pairing gap predominantly opens at $\pm Q_{i}$'s shown in Fig.\ref{fig:FS and valley-crossing points}, it is convenient to describe pairing %function $\Delta_{K\bar K}$ and 
in terms of the electron dispersion within the patches around $\pm Q_{i}$'s, treating $\pm Q_i$ as a patch index. Namely, we define a gap function near $\pm Q_i$ as $\Delta_{\vec K\bar {\vec K};\pm Q_{i}}(\vec k)$, where $\vec k$ is measured from $Q_i$, and $k'$ is measured from $-Q_i$. Here, $\vec k,\vec k'\ll k_F$. Accordingly, we model the electron energy near $\pm Q_i$ as:
\be\label{eq:epsilon}
\epsilon_{\pm i,\vec k} = v_F \vec{n}_{\pm i} \cdot %\lp
\vec k
%-\vec Q_{i}\rp
+ |\vec{n}_{\pm i} \times %\lp 
\vec k
%-\vec Q_{i}\rp
|^2/2m_\perp
\ee
where $\vec n_{\pm i}$ are the unit vectors  normal to the Fermi surface at $\pm Q_{i}$. Applying this model to describe pairing and keeping only the scattering processes in which an electron is scattered from a patch near $Q_i$ to a patch near $-Q_i$, which is the most singular contribution, we find
\[ %\label{eq:self-consistency main text}
\Delta_{\vec K\bar {\vec K};Q_{i}}(\vec k,\omega) = -\sum_{\vec k'\omega'} \frac{\Gamma_{\omega+\omega',\vec k+\vec k'}\Delta_{\vec K\bar {\vec K};-Q_{i}}(\vec k',\omega')}{\omega'^2 + \epsilon^2_{-i, \vec k'}}, 
\]
The analysis of this equation is detailed in \cite{SM}. Below we describe the main predictions. 

%First, 
Since $\Gamma_{\omega+\omega',\vec k+\vec k'}$ is positive, the gap equation %self-consistency equation, \eqref{eq:self-consistency main text}, 
predicts a sign-changing solution $\Delta_{\vec K\bar {\vec K};Q_{i}}=-\Delta_{\vec K\bar {\vec K};-Q_{i}}$(see Sec.\ref{sec:leading channel} in Ref.\cite{SM}). 
This yields two degenerate pairing channels that respect the symmetry group (see Sec.\ref{sec:symmetry} in Ref.\cite{SM}). These are the p-wave $p_x\pm ip_y$ channels identical to the ones identified 
for RTG \cite{chatterjee2022inter} and moir\'e graphene \cite{wang2021topological}. %We note that the 
A linear superposition of these two channels gives rise to a $p$-wave channel which breaks the rotational symmetry, which has a $T_{\rm c}$ degenerate with that of $p_x\pm ip_y$ channels. In a  recent experiment\cite{holleis2023ising} in WSe$_2$-supported BBG samples the SC phases are found to emerge on top of, or next to, a nematic phase where three-fold rotation symmetry is spontaneously broken. This suggests that the $p$-wave superconductivity wins in these systems.

An interesting behavior of SC that is unique to the three-pocket Fermi sea is an increase in $T_{\rm c}$ near the termination of SC phase. Indeed, %according to our discussion above, under the termination condition, 
upon the appearance of ghost crossings the Fermi pockets in two valleys become nearly tangential at the crossing point. In this case, an approximate nesting at the ${\vec K}$-$\bar {\vec K}$ pocket crossings by a  vector $2\vec K$ allows pairing to occur on a larger Fermi surface segment near the crossing points, which leads to an enhancement in superconducting $T_{\rm c}$.  This behavior is manifest %can be directly read off from 
in the expression for $T_{\rm c}$ derived in \cite{SM},  Sec.\ref{sec:Tc}:
\be \label{eq:Tc main text}
T_{\rm c} = 2\omega_0 e^{-\frac{1}{\lambda}}, \quad \lambda = \frac{U}{8v_Fl_0\delta\sin (\theta/2) }, \quad \omega_0 = \frac{\delta v_F}{l_0}.
\ee
with $\theta$ the angle between the ${\vec K}$ and $\bar {\vec K}$ Fermi surfaces at the crossing points (see Fig.\ref{fig:FS and valley-crossing points} c)). The increase in $T_{\rm c}$ occurs because $\theta$ vanishes when the Fermi surfaces %in ${\vec K}$ and $\bar {\vec K}$ are 
become tangential. We expect  the divergence of $\lambda$ at $\theta\to 0$ to be cut off by the dispersion curvature described by the quantity $m_\perp$ in  \eqref{eq:epsilon}); this effect %does not show up explicitly
is not manifest in \eqref{eq:Tc main text} as it is subleading for finite $\theta$. However, $m_\perp$ will limit the phase volume in $\vec k$-space where pairing can occur when $\theta\rightarrow 0$, thereby, cutting off the divergence of $\lambda$ and $T_{\rm c}$. The enhancement in $T_{\rm c}$ near the termination point probably explains why IVC fluctuations dominate the pairing despite %the potential existence 
possible presence of other fluctuations, e.g. due to nematic order conjectured in Ref.\cite{holleis2023ising}.
%fluctuation which is conjectured in WSe$_2$-supported BBG. 

Unfortunately, the existing data are insufficient to map out this interesting behavior, though % near the superconducting phase termination point. 
it is somewhat consistent with the superconducting phase in Fig.\ref{fig:BBG+WSe2_all} widening near the termination point. %, however in itself this does not provide a strong evidence. 
Verifying the predicted nonmonotonic behavior in $T_{\rm c}$ near the termination point is an interesting direction for future experiments.

Next, we use a realistic bandstructure %of BBG and WSe$_2$-supported BBG 
to obtain a condition for the ghost Fermi surface crossings to exist and demonstrate an agreement with the observed onset of superconductivity. 
We first present the analysis for WSe$_2$-supported BBG. % using the measured data 
In Ref.\cite{holleis2023ising} two superconductivity phases were found. Here, we  focus on the SC$_1$ phase, which emerges from an isospin-unpolarized parent state. %One can generalize the analysis to 
%We leave the analysis of the SC$_2$ phase for future work since 
The phase SC$_2$ emerges from a parent state with a pocket polarization\cite{holleis2023ising} that needs an analysis that accounts for the interaction effects, which we leave to future work. 
%, which is more complicated and requires a more complicated analysis that properly accounted for interaction effects. 

We predict the onset of valley-crossings by numerically calculating the single-particle band dispersion in the isospin-unpolarized phase. We model the single-particle in WSe$_2$-supported BBG using the Hamiltonian
\be\label{eq:BBG+WSe2}
H = H_{\rm BBG}+ H_{\rm SOI}
.
\ee
The first term $H_{\rm{BBG}}$ is the four-band tight-binding model % (see Sec.\ref{sec:model} in \cite{SM}),
given in the basis $\left\lbrace c_{A,1}^{\eta,s},c_{B,1}^{\eta,s},c_{A,2}^{\eta,s},c_{B,2}^{\eta,s}\right\rbrace $ (with A and B the sublattice indices, 1 and 2 the layer indices, $\eta=\pm1$ a valley label % ${\vec K}$, $\bar {\vec K}$, 
and $s$ the spin index) \cite{mccann2013electronic,jung2014accurate} and \cite{SM} Sec.\ref{sec:model}:
\be\label{eq:H_BBG}
H_{\rm{\rm BBG}} = \lp
\begin{matrix}
	u/2             & v \pi^\dagger & -v_4\pi^\dagger & v_3\pi  \\
	v\pi            & u/2+\Delta'   & t_1             & -v_4\pi^\dagger \\
	-v_4\pi         & t_1           & -u/2+\Delta'    &  v\pi^\dagger   \\
	v_3\pi^\dagger  & -v_4\pi       & v\pi            & -u/2
\end{matrix}
\rp
\ee
where $\pi = \hbar \lp \eta k_x+ik_y \rp$, %$\eta=\pm1$  corresponding to valley ${\vec K}$ and $\bar {\vec K}$ respectively, 
$k_x$ and $k_y$ are the $x$ and $y$ components of momentum measured from ${\vec K}$ or $\bar {\vec K}$. The quantity $u$ is the interlayer bias, % between the upper and lower layers.
%This Hamiltonian gives a trigonal-warped conduction band \cite{Jung2011lattice}
$t_1$ is the interlayer hopping parameter, $v$, $v_3$, $v_4$ are associated with  microscopic hopping amplitudes of the values given in \cite{jung2014accurate}. %Here 
%For these quantities, we use the %nominally 
%realistic values given in Ref.\cite{Jung2011lattice}.} % for these hopping amplitudes.
The second term $H_{\rm{SOI}}$, \eqref{eq:BBG+WSe2}, represents an Ising spin-orbital interaction (SOI) induced by the proximate WSe$_2$ layer, which takes the valley Ising form\cite{zhang2022spin}:
\be
H_{\rm{\rm SOI}} = \lambda_{\rm I} \eta \sigma_z
\ee
where $\eta=\pm1$ for valley ${\vec K}$ and $\bar {\vec K}$, $\sigma_z$ is the Pauli matrix for spin in the out-of-plane direction.

To determine how the onset of Fermi surface crossings compares with measured SC phases, two parameters in model \eqref{eq:BBG+WSe2} must be obtained by careful analysis of existing data.
%For comparing the onset of the  Fermi surface crossings with the measured SC phases, there are two parameters in the model \eqref{eq:BBG+WSe2} that can be determined by a careful analysis of the existing data. 
One is the interlayer bias $u$, which is proportional to the transverse electric field $D$, yet the ratio between $u$ and $D$ in general  is not exactly known (see discussion below). Another is the spin-orbit coupling $\lambda_{\rm I}$. %Although there are experiments that measured $\lambda_{\rm I}$ \cite{holleis2023ising}, we cannot directly use those values here. The reason will be discussed shortly.

We determine these two quantities using the quantum oscillations measured in Ref.\cite{holleis2023ising}. This measurement accurately gives the carrier densities where two distinct transitions of Fermi surface topology occur in the minority isospin species ${\vec K}\downarrow$ and $\bar{\vec K}\uparrow$. One is the transition from a single Fermi sea to an annular Fermi sea occurs at $n=9.9\times 10^{11}\rm{cm}^{-2}$. The other is the transition from the annular Fermi sea to a three-pocket Fermi sea occurs at $n=9.7\times 10^{11}\rm{cm}^{-2}$. Using these two data points as constraints, we are able to determine the numerical values of the two unknown parameters: 
%$\frac{u}{\rm{meV}} =  \frac{(0.047\pm0.001) D}{\rm{mV/nm}}$ and $\lambda_{\rm I} = (7\pm 1)\rm{meV}$. 
\[
\frac{u}{\rm{meV}} =  \frac{(0.047\pm0.001) eD}{\rm{mV/nm}}
,
\quad \lambda_{\rm I} = (7\pm 1)\rm{meV}
.
\]
Using these values, %we follow the same procedure as in the analysis of BBG, 
we study the evolution of Fermi seas % of the minority isospin species inside 
within the symmetry-unbroken ``vanilla'' phase. We focus on the Fermi seas of the minority isospin species %$\vec K\downarrow$ and $\bar{\vec K}\uparrow$ 
as the majority isospin species feature a single Fermi sea that does not experience any qualitative change in this regime. 
In the regime where SC${}_1$ phase occurs, the majority species feature valley crossings of the $\vec K$ and $\bar{\vec K}$ Fermi seas with an order-one angle $\theta\sim O(1)$ at the crossing and, consequently, no enhancement in superconductivity due to small $\theta$ values similar to the one demonstrated by the minority species. We therefore focus on minority species, deferring the analysis of majority species till later. 
% \addQ{The possibility of pairing in majority isospin will be discussed below.} 
We determine the transition from overlapping to non-overlapping Fermi seas in minority species, finding a transition marked by the yellow lines in Fig.\ref{fig:BBG+WSe2_all}. The solid yellow line lies inside the symmetry-unbroken ``vanilla'' phase where a single-particle band calculation can be trusted. The dashed yellow line %is merely a guide to the eye. It does not have a physical meaning since it enters 
drawn across the ordered phase where a single-particle band calculation is invalid  merely provides a guide to the eye. Notably, the yellow line crossing with the phase boundary agrees well with the SC phase onset.
%We find that the termination point of SC phase extracted from measured data (the transition point from cyan to black line) coincides with the occurrence of valley crossing predicted by calculation. 
This provides strong evidence for the pairing governed by a back-scattering mechanism, such as the IVC QCP scenario.

%We briefly comment on 
We note that the value of $\lambda_{\rm I}$ extracted and used above is a few times greater %: This value, which is several times larger as compared to the 
than the value $\lambda^{(0)}_{\rm I}=1.6$\,meV inferred from measurements in a strong out-of-plane $B$ field \cite{holleis2023ising}. We believe that this discrepancy is reasonable. Indeed, the value of $\lambda_{\rm I}$ that should be plugged in our simulation is not the bare SOI strength, but rather the effective interaction renormalized by strong interactions in a flat-bottom BBG band. The vertex corrections that govern this renormalization are expected to be large since the system is in the regime close to all kinds of spin and valley Stoner instabilities. 
%strength of the effective one, which involves the vertex correction due to strong interaction in the BBG's flat band bottom. This vertex correction is expected to be large since the system is in the regime close to all kinds of spin and valley Stoner instabilities. 
In contrast, $\lambda^{(0)}_{I}$ measured in a large $B$ field\cite{holleis2023ising} is largely insensitive to this physics, which explains the above disparity in $\lambda_{I}$ values. %breaks down the original BBG band structure.

Next, we turn to the BBG/hBN case\cite{zhou2022isospin}. We continue to use the four-band model given in \eqref{eq:H_BBG}. %, with the parameters given in Ref.\cite{Jung2011lattice}. 
Here, however, unlike the case of WSe$_2$-supported BBG, the relation between the interlayer bias parameter $u$ and the experimental displacement field $D$ is not accurately known. %Unlike the case of WSe$_2$-supported BBG, 
For BBG/hBN quantum oscillation measurements\cite{zhou2022isospin} do not provide sufficient information to extract the ratio between interlayer bias $u$ and displacement field $D$. However they do give useful upper and lower bounds for $u/D$. 

Namely, quantum oscillation measurements \cite{zhou2022isospin} reveal that: 1) the isospin-unpolarized Sym$_4$ phase below the isospin-polarized PIP$_2$ phase (shown in Fig. \ref{fig:BBG_all} b) has a single Fermi surface per isospin; 2) the isospin-unpolarized Sym${_{12}}$ phase above PIP$_2$ features three distinct pockets per isospin.
%Namely, the quantum oscillations measurements in Ref.\cite{zhou2022isospin} indicate that: 1) the isospin-unpolarized phase Sym$_4$ beneath the isospin-polarized phase PIP$_2$ [see Fig.\ref{fig:BBG_all} b)] has a single Fermi surface for each isospin; 2) in the isospin-unpolarized phase Sym$_{12}$ above the PIP$_2$ phase the Fermi surface consists of three distinct pockets for each isospin. 
%
%on the \textcolor{red}{right}
%%left ????
%of the isospin polarized phase (PIP$_2$) is Sym$_{12}$ \addLL{[the phase above the black curve in Fig.\ref{fig:BBG_all} b)]} where Fermi sea is always three-pocket .  
%
We find that in order to reproduce these two observations, the value $u/D$ should fall in the range: $0.057 < \frac{u }{eD \cdot \rm{nm}} < 0.072$. Accordingly, as a best guess, we pick a value in the middle of this window, $\frac{u }{eD \cdot \rm{nm}}=0.065$. 
% Using this value, we obtain the transition line between overlapping and non-overlapping Fermi seas (shown as the yellow line in Fig.\ref{fig:BBG_all}) that agrees very well with the SC phase termination point.
With this value, we derive the transition line between overlapping and non-overlapping Fermi seas, indicated as the yellow line in Fig.\ref{fig:BBG_all}, which closely matches the  %termination 
emergence point of the SC phase.

It is worth noting that several experiments have attempted to measure the $u/D$ ratio for BBG, yielding vastly different values that do not fall within the range inferred from our fermiology analysis. % mentioned earlier. 
%We note parenthetically that several experiments attempted to measure the $u/D$ ratio for BBG arriving at very different values, none of which falls in the range inferred above from the fermiology analysis. 
Namely, Refs.\cite{zhang2009direct} and \cite{zhang2022promotion} find $\frac{u }{eD \cdot \rm{nm}}  =  0.1$ and $  0.033$, respectively. 
%Given the large disparity in these values, probably reflecting  differences in electrostatic environments of different devices, we avoided using them directly. Instead, the parameter $u/D$ was picked as described above so that the fermiology it predicts maximally resembles that inferred from the measurements.
Due to significant variations in values, likely due to electrostatic differences in devices, we %didn't use 
refrained from using them directly. Instead, we selected $u/D$ as described above to predict fermiology that best matches the measurements.

%\sout{In conclusion,} \addQ{[\bf There is another "in conclusion" in line 405, so I reword it:]} \sout{\addQ{The analysis above shows that}}
%Using these values to predict the boundary at which the ghost valley crossings emerge we obtain the yellow lines shown in Figs.\ref{fig:FS and valley-crossing points} and \ref{fig:BBG_all}. 
To restate the main result of our analysis, 
in both BBG and in BBG/WSe${}_2$ the lines that mark the emergence of ghost Fermi surface crossings match perfectly the points of the onset of superconductivity. As discussed above, the emergence of valley crossings strongly impacts the pair scattering kinematics, favoring backscattering. To the contrary, it has little impact on the density of states at the Fermi level %nor it affects 
or the e-e interaction strength. Therefore the observed behavior is difficult to understand within a conventional BCS superconductivity framework but is naturally explained by the IVC superconductivity mechanism. 
%\sout{\addQ{This is the main result of this paper.}} \addQ{\bf [Trying to highlight this claim because the discussion below is long and, therefore, might distract readers and cause questions. I do not want to shorten the discussion below. I only want to remind readers that the result above is our main claim.]}

%\addQ{We note parenthetically that} 
This conclusion is further supported by the nonmonotonic behavior of superconductivity near the onset (a rise followed by a drop) observed in BBG/WSe$_2$. This observation is explained by the enhancement of pairing at small angles $\theta$ between Fermi surfaces at the ghost crossing discussed above. It is interesting to compare the nonmonotonic behavior of superconductivity in BBG/WSe$_2$ with the monotonic behavior observed at superconductivity onset in BBG/hBN samples. 
%LL It is interesting to compare the superconducting (SC) behavior in WSe$_2$-supported BBG with that in BBG/hBN samples. In the former case, as discussed, the SC phase is slightly widened near the onset of valley crossings, implying an enhancement. In contrast, in the BBG/hBN case (see Fig. \ref{fig:BBG_all}), the SC phase monotonically narrows as the onset of valley crossings is approached. 
We believe that this difference can be attributed to the constraints imposed by isospin orders and Fermiology. Specifically, the SC phase in BBG/hBN must lie outside the PIP$_2$ phase which is not necessarily compatible with the SC order, and below the boundary where valley crossings occur (yellow line in Fig. \ref{fig:BBG_all}). These constraints limit the SC phase to a narrow wedge in the phase diagram (Fig. \ref{fig:BBG_all}), preventing the enhancement of the SC phase at the onset of valley crossings. In comparison, in the BBG/WSe$_2$ these constraints are lifted. Since the  ``vanilla'' phase lies below the phase boundary, the onset of valley crossings (solid yellow line in Fig. \ref{fig:BBG+WSe2_all}) extends downwards and therefore does not constrain the SC phase. %allowing it to 
%which extend upwards (cyan region in Fig. \ref{fig:BBG+WSe2_all}).

%LL {\bf for general readers:}
%LL \addQ{This comparison of two SC systems provides insight into a general way of enhancing back-scattering type SC. Specifically, to achieve this goal, we generally need to engineer the band structure or interaction in such a way that the onset line of valley crossings and the ordering phase boundary leave a sufficiently wide phase space for SC to benefit from the enhancement near the onset. As suggested by the WSe$_2$-supported and BBG/hBN cases, where the difference in $T_{\rm c}$ is on the order of 10, this enhancement can be significant.}
%\addQ{From this comparison of two SC systems we learn a general way of enhancing these back-scattering type SC. Namely, to achieve this goal, we generally needs to engineer the band structure or the interaction so that the onset line of valley-crossings and ordering phase boundary leave a sufficiently wide phase space for SC to freely enjoy the enhancement near the onset. As implied by the WSe$_2$-supported case and BBG/hBN case where the difference of Tc is $O(10)$, this enhancement can be generally significant.} %This is align with the WSe$_2$-supported case, and is unlike the case of BBG/hBN where these two lines form a sharp angle.

%{\bf for graphene people:}
Lastly, we believe that the IVC pairing revealed by our analysis is generally applicable to other observed SC phases, such as the SC$_2$ phase in BBG/WSe${}_2$. %We did not analyze it here because that 
Here a conclusive analysis would require more knowledge of the isospin phase diagram, which is currently being investigated by several groups \cite{xie2023flavor, wang2023electrical}. Nonetheless, the SC$_2$ phase, which is a wedge embedded between different isospin orders, shows an abrupt onset which is likely related to a ghost valley crossing %, as it also shows an abrupt termination point in the phase diagram 
(see Fig. \ref{fig:BBG+WSe2_all}).
%The only difficulty is that 

%but we do not directly use one of them. This is because of two reasons: a) the measured $u/D$ is different for different experiments: For example, the experiment\cite{zhang2009direct} finds a mapping relation: $\frac{u }{eD \cdot \rm{nm}}  =  0.1$, whereas the experiment\cite{zhang2021ascendance} finds a quite different one: $\frac{u }{eD \cdot \rm{nm}}  =  0.033$. b) None of the two measured values falls into the required regime discussed above. In other words, if use any one of them directly in our model, then it does not correctly reproduce the fermiology seen in the experiment. Since the fermiology is so important for the main physics discussed here, we have to choose the parameter $u/D$ so that the fermiology it predicts maximally resembles the measured ones.

In conclusion, the sudden appearance of SC phases coincides with the appearance of the ${\vec K}$-$\bar {\vec K}$ ghost Fermi surface crossings in both BBG and %SC$_1$ phase 
WSe$_2$-supported BBG. %This signature provides strong evidence for pairing in both systems driven by quantum-critical fluctuations that produce pairing interaction of a backscattering type. 
This behavior suggests that quantum-critical fluctuations drive the pairing in both systems, favoring a backscattering-type pairing interaction due to the IVC order as the glue for superconductivity over other candidates like valley-polarization order \cite{dong2022spin}. 
%This singles out a backscattering-type pairing interaction originating from the IVC order as a likely pairing glue for superconductivity, favoring it over other candidates such as the valley-polarization order \cite{dong2022spin}. 
%This behavior 
Overall, it is not compatible with conventional phonon mechanisms\cite{chou2021acoustic,chou2022acoustic}, nor with the conventional Kohn-Luttinger mechanisms\cite{cea2022superconductivity,cea2023superconductivity}, pointing to % strong evidence for 
a mechanism that involves a soft quantum-critical mode as a pairing glue. 
Last but not least, it highlights tuning bands through ghost crossings as an attractive pathway to enhance superconductivity. 
%However, we note that other order types may exist that drive pairing of a backscattering type, e.g. nematic order in WSe$_2$-supported BBG.\textcolor{r}{I dont understand what we think is fluctuating here.  we need to discuss this point over zoom. } Nevertheless, despite some ambiguity in details, our
%Our analysis clearly indicates that superconductivity in these systems is driven by a mechanism in which pair interaction is dominated by backscattering.  

%\addQ{We highlight that the signature studied above is a generic hallmark of the Cooper pair kinematics that is readily applicable to many tunable systems, such as Sr$_2$RuO$_4$ where the van Hove singularity can be tunned by strain and iron-based superconductors where strain is used to adjust the competition between nematic order and superconductivity. Our analysis can also be generalized to the systems where the ratio between kinetic and interaction energy is tunable through applying pressure, for example, CeCoIn$_5$, $\kappa$-phase organic superconductors, and heavy fermion systems such as UPt$_3$. Searching for the signature of Cooper pair kinematics will potentially reveal new clues for the mechanism of superconductivity in these systems.}

We thank A. F. Young and S. Nadj-Perge for sharing unpublished data, and A. V. Chubukov and J. G. Analytis for fruitful discussions. 
This work was supported by the Science and Technology Center for Integrated Quantum Materials, National Science Foundation Grant No. DMR1231319, and Army Research Office Grant No. W911NF-18-1-0116. 
P. L. acknowledges the support by DOE office of Basic Sciences Grant No. DE-FG02-03ER46076.

\bibliography{ref}

\begin{appendix}

\begin{widetext}

\section{Model}\label{sec:model}
In this section we provide details on the model used in the main text. %As we will see, the model we used is standard: 
Our  single-particle Hamiltonian is identical to that in Ref.\cite{mccann2013electronic} whereas the interaction model is same as that in Ref.\cite{dong2022spin}.

The electron single-particle Hamiltonian for BBG is a $4\times 4$ matrix expressed in the basis of $\left\lbrace c_{A,1}^{\eta,s},c_{B,1}^{\eta,s},c_{A,2}^{\eta,s},c_{B,2}^{\eta,s}\right\rbrace $ (A and B are sublattice indices, 1 and 2 are the layer indices, $\eta=\pm1$ labels valley ${\vec K}$ and $\bar {\vec K}$, and $s$ is the spin index)  \cite{mccann2013electronic}:
\be\label{eq:H_el^0}
H_{\rm BBG} = \lp
\begin{matrix}
	u/2             & v \pi^\dagger & -v_4\pi^\dagger & v_3\pi  \\
	v\pi            & u/2+\Delta'   & t_1             & -v_4\pi^\dagger \\
	-v_4\pi         & t_1           & -u/2+\Delta'    &  v\pi^\dagger   \\
	v_3\pi^\dagger  & -v_4\pi       & v\pi            & -u/2
\end{matrix}
\rp
\ee
where $\pi = \hbar \lp \eta k_x+ik_y \rp$,	$k_x$ and $k_y$ are the $x$ and $y$ components of momentum measured from ${\vec K}$ or $\bar {\vec K}$. This Hamiltonian gives a trigonally-warped conduction band. In the isospin-polarized phase on which the SC emerges, quantum oscillations measurement shows that the Fermi sea in each isospin consists of three pockets, as shown in panels (ii) and (iii) of Fig.\ref{fig:FS and valley-crossing points} c) in main text.

The electron-electron interaction in BBG is modeled as an intervalley local interaction
\be\label{eq:H_int^0}
H_{int}^{(0)} = \frac12 U\sum_{l,\alpha,q,\tau, s}:\rho^{(0)}_{q,\alpha,l,\tau, s}\rho^{(0)}_{-q,\alpha',l',\tau',s'}:
\ee
where $\rho^{(0)}_{q,\alpha,l,\eta,s} = \sum_{\vec k}c^{\dagger}_{\vec k+\vec q,\alpha,l,\eta, s}c_{\vec k,\alpha,l,\eta', s'}$. In the analysis below we project $H_{int}^{(0)}$ onto the conduction band which is the only band relevant for SC. In the regime of $D$ much larger than all other elements in Eq.\eqref{eq:H_el^0}, the projected interaction Hamiltonian is approximately given by
\be\label{eq:Hint}
H_{int} = \frac12 U\sum_{q,\tau \tau' s s'}:\rho_{q,\eta,s}\rho_{-q,\eta',s'}:,\quad 
\ee
Here the quantities $\rho_{q,\eta,s}$ are projected density operators defined as $\rho_{q,\tau,s} = \sum_{\vec k}\psi^{\dagger}_{\vec k+\vec q,\tau,s}\psi_{\vec k,\tau,s}$, where $\psi_{\vec k,\tau,s}$ is an electron operator projected onto the conduction band.

\section{The scattering vertex function near IVC instability}
\label{sec:pairing interaction}
In this section, we consider the Cooper pair scattering vertex function mediated by the soft mode describing IVC order fluctuations in the normal state. This quantity is overall similar to that considered in Ref.\cite{chatterjee2022inter,wang2021topological}. However, unlike in Ref.\cite{wang2021topological} which directly starts from a spin-valley-fermion model, here we derive it from microscopic model Eq.\eqref{eq:H_BBG} and Eq.\eqref{eq:Hint} explicitly. Our derivation below does not account for the intervalley exchange interaction considered in Ref.\cite{chatterjee2022inter}. As we will see, the frequency dependence of the Cooper pair scattering vertex function slightly differs from that considered in Ref.\cite{wang2021topological,chatterjee2022inter}.  

We start by defining the IVC order parameter as follows\cite{dong2021}
\be
O_{\rm IVC} = \sum_{s,\vec k} \langle \psi^\dagger_{\vec k,\eta,s}\tau_{x,\eta\eta'} \psi_{\vec k,\eta',s} \rangle
\ee
here $\tau_x$ is the Pauli matrix in $x$ direction. Near the onset of this order, the most divergent contribution to vertex function is dominated by the ladder diagram shown in main text Fig.\ref{fig:FS and valley-crossing points} a), in which a Cooper pair at momenta ${\vec K}+\vec k$ and $\bar {\vec K}-\vec k$ is scattered to a pair at ${\vec K}+\vec k'$ and $\bar {\vec K}-\vec k'$. The scattering vertex given by this diagram is clearly a function of the total frequency $\omega+\omega'$ and the total momentum $\vec k+\vec k'$ and takes the following form:
\begin{align} \label{eq:Gamma_1}
	&\Gamma_{\omega+\omega',\vec k+\vec k'}= \frac{U}{1+U\Pi_{{\vec K}\bar {\vec K}}(\omega+\omega',\vec k+\vec k')},\\
	&\Pi_{{\vec K}\bar {\vec K}}(\nu, \vec q) = \sum_{\omega,\vec p} G_K(\omega+\nu, \vec k+\vec q) G_{\bar {\vec K}}(\omega,\vec k)  
\end{align}
This interaction diverges for $\omega+\omega'=0, \vec k+\vec k'=0$ because the Stoner criterion in IVC channel is given by\cite{dong2021superconductivity}
\be\label{eq:Stoner}
1+U\Pi_{{\vec K}\bar {\vec K}}(0,0) \rightarrow 0
\ee

Distinct from the valley-polarization susceptibility which contains a frequency-dependent term singular in momentum $\frac{|\nu|}{|q|}$ due to Landau damping, the intervalley particle-hole susceptibility does not contain such a term. This is because the inter-valley zero-frequency momentum-${\vec K}-{\bar {\vec K}}$ particle-hole excitations have a measure-zero phase space. Namely, in our case, the $q=0$ inter-valley particle-hole gap closes only at six points: $\vec k=\pm \vec Q_i$, ($i=1,2,3$). This is distinct from usual particle-hole excitations in Fermi liquid where the $q=0$ particle-hole excitation is gapless for any $\vec k$ on the Fermi surface.

To see this explicitly, we calculate the frequency and momentum dependence inter-valley  particle-hole susceptibility below %in the following section \ref{sec:calculate Pi_KK'}
and find
\be\label{eq:Pi(nu,q)}
\Pi_{{\vec K}\bar {\vec K}}(i\nu,q) =  \Pi_{{\vec K}\bar {\vec K}}(0,0) + \frac{3|\nu|}{4\pi v_F^2\sin\theta}  + \alpha q^2. 
\ee
Here $\alpha$ is a parameter depending on band dispersion details. $\theta$ is the angle between the fermi surface in valley ${\vec K}$ and that in valley $\bar {\vec K}$ at the valley-crossing points. In the analysis below, we take $\sin\theta \sim O(1)$ unless specified otherwise.

%============
%\section{Calculate intervalley particle-hole susceptibility $\Pi_{{\vec K}\bar {\vec K}}(i\nu,q)$ }\label{sec:calculate Pi_KK'} 
%In this section, we explicitly derive  the result we used in Eq.\eqref{eq:Pi(nu,q)}, which is the frequency and momentum dependence of IVC susceptibility. 
To see this explicitly, below we pause and derive the frequency and momentum dependence of IVC susceptibility Eq.\eqref{eq:Pi(nu,q)}. As we will see, the susceptibility does not contain a Landau damping term $|\nu|/|q|$ in the denominator. This is quite different from the textbook spin-susceptibility % discussed in textbook 
(e.g., see Ref.\cite{Coleman2015}) or the valley-polarization susceptibility discussed in Ref.\cite{dong2022spin} where Landau damping term shows up.

%For simplicity, in this section we will set $\delta=0$.
\begin{align}
	\Pi_{{\vec K}\bar {\vec K}}(i\nu,q) &=  \sum_{\omega,p} G_{\vec K} (\omega+\nu,\vec k+\vec q) G_{\bar {\vec K}} (\omega,\vec k) 
	\\&= \sum_{\vec k,\omega} \frac{f_{\vec k+}-f_{\vec k-}}{\epsilon_{\vec k+}-\epsilon_{\vec k-}-i\nu}, 
\end{align}
where $\vec k_\pm = \vec k\pm \frac{\vec q}{2} \pm {\vec K}$. As a reminder, $\bar {\vec K}=-{\vec K}$.

\begin{align}
	\Pi_{{\vec K}\bar {\vec K}}(i\nu,0) &=  \sum_{\vec k} \frac{\lp f_{{\vec K}+\vec k}-f_{\bar {\vec K}+\vec k}\rp \lp\epsilon_{{\vec K}+\vec k}-\epsilon_{\bar {\vec K}+\vec k}+i\nu\rp}{(\epsilon_{{\vec K}+\vec k}-\epsilon_{\bar {\vec K}+k})^2 + \nu^2 }\\
	& = 6 \sum_{k} \frac{\lp f_{{\vec K}+\vec Q+\vec k}-f_{\bar {\vec K}+\vec Q+ \vec k}\rp \lp \Delta \vec v \cdot \vec k +i\nu\rp}{\lp \Delta \vec v \cdot \vec k \rp^2 + \nu^2 }, \quad \Delta \vec v = \vec v_{{\vec K}+\vec Q}-\vec v_{\bar {\vec K}+\vec Q} \\
	& = -6 \int_{-k_F}^{k_F} \frac{dk_x}{2\pi} \frac{2k_x\sin\theta}{2\pi} \frac{  2v_F k_x \sin\theta }{\lp 2v_F k_x \sin\theta \rp^2 + \nu^2 } . 
\end{align}
Here $\theta$ is the angle between the fermi surface in valley ${\vec K}$ and that in valley $\bar {\vec K}$ at the valley-crossing points. To evaluate this integral properly, we calculate the difference between the dynamical polarization function $\Pi_{{\vec K}\bar {\vec K}}(i\nu,0) $ and the static polarization function $\Pi_{{\vec K}\bar {\vec K}}(0,0)$:
\be
\Pi_{{\vec K}\bar {\vec K}}(i\nu,0)- \Pi_{{\vec K}\bar {\vec K}}(0,0)   = 6 \int_{-k_F}^{k_F} \frac{dk_x}{(2\pi)^2 v_F} \frac{  \nu^2 }{\lp 2v_F k_x \sin\theta \rp^2 + \nu^2 } = \frac{3|\nu|}{4\pi v_F^2\sin\theta} . \label{eq:Pi(nu,0)}
\ee
This result is the leading order frequency dependence in $\Pi_{{\vec K}\bar {\vec K}}(i\nu,0)$. 

Next, we determine the form of leading order momentum dependence from symmetry. We know from $C_3$ rotation symmetry that the momentum-dependent part of intervalley particle-hole susceptibility $\Pi_{{\vec K}\bar {\vec K}}(0,\vec q)$ has to be a function of $q^2$. Therefore, for small $q$ ($q\ll k_F$),
\be \label{eq:Pi(0,q)}
\Pi_{{\vec K}\bar {\vec K}}(0,\vec q) =  \Pi_{{\vec K}\bar {\vec K}}(0,0) + \alpha q^2. 
\ee
Here $\alpha$ is a parameter depending on band dispersion details. Then, accounting for both frequency dependence Eq.\eqref{eq:Pi(nu,0)} and momentum dependence Eq.\eqref{eq:Pi(0,q)}, we arrive at the expression of intervalley susceptibility $\Pi_{{\vec K}\bar {\vec K}}(i\nu,q)$ given in Eq.\eqref{eq:Pi(nu,q)}.
%===================

With the results of IVC susceptibility Eq.\eqref{eq:Pi(nu,q)} derived above, we proceed to derive the pairing interaction. Plugging Eq.\eqref{eq:Pi(nu,q)} and Eq.\eqref{eq:Stoner} into Eq.\eqref{eq:Gamma_1} we find the pairing interaction $\Gamma_{kk'}$ that takes the following form given in main text Eq.\eqref{eq:Gamma main text} 
\be\label{eq:Gamma}
\Gamma_{\omega \vec k, \omega' \vec k'} 
=  \frac{U}{ \kappa |\omega+\omega'| + l_0^2 (\vec k+ \vec k')^2 +\delta^2}, 
\ee
with the parameter $\kappa = \frac{3U}{4\pi v_F^2\sin\theta}$ and the lengthscale $\l_0$ is defined as $l_0 = U\alpha$. For simplicity, here we take $l_0=1/k_F$ since it is the only lengthscale in this problem. The quantity $\delta$ describes the ``distance'' from  the IVC quantum criticality, it is defined as $\delta = \sqrt{1+U\Pi_{{\vec K}\bar {\vec K}}(0,0)}$. Eq.\eqref{eq:Stoner} indicates that $\delta\rightarrow 0$ near the IVC instability. 
%For simplicity, below we will set $\delta=0$.

This scattering amplitude has a positive sign, which is naively pair-breaking. However, unlike the VP-mediated interaction which is divergent at a small angle scattering $\vec k'=\vec k$, the strength of this interaction is maximized for backward scattering $k'=-k$ since $\delta\rightarrow 0$. This indicates an attractive pairing interaction in the non-s-wave channel. 

This scenario resembles the spin-fluctuation-mediated near a ferromagnetic quantum critical point\cite{fay1980coexistence,wang2001superconductivity}, where spin fluctuation generates a repulsive backward scattering, leading to an effective pairing interaction (attraction) in the $p$-wave channel. However, there is some difference in our case, since enhanced backward scattering here is intravalley scattering ${\vec K}+\vec p\rightarrow {\vec K}-\vec p$, rather than $\vec p\rightarrow -\vec p$ in usual metals. Since ${\vec K}+\vec p$ is not associated with ${\vec K}-\vec p$ by symmetry operation (see discussion below), they cannot, in general, simultaneously get close to the Fermi surface. This suggests that such ${\vec K}+\vec p\rightarrow {\vec K}-\vec p$ scattering does not always contribute to pairing, except for some special momentum $p$. Below, we will discuss this in detail.

\section{Symmetry analysis of the superconducting gap function} \label{sec:symmetry}
In this section, we describe the symmetry of the pairing gap function $\Delta_{\vec K\bar{\vec K}}$. This symmetry analysis, which is essentially the same as that in Ref.\cite{chatterjee2022inter}, is required for the classification of different pairing channels and for identifying the leading pairing channel. % discussed below.

To start, we describe the superconductivity in BBG using the following BCS Hamiltonian:
\be
H_{BCS} = \lp
\begin{matrix}
	\epsilon_{{\vec K}+\vec k} & \Delta_{{\vec K}\bar {\vec K}} \\
	\overline{\Delta}_{{\vec K}\bar {\vec K}} &  -\epsilon_{\bar {\vec K}-\vec k} 
\end{matrix}
\rp
\ee
where $\Delta_{{\vec K}\bar {\vec K}}$ represents the anomalous pairing vertex. Here we have suppressed the physical spin indices $\uparrow$/$\downarrow$ since we are focusing on the parallel spin pairing, i.e. either $\uparrow\uparrow$ pair or $\downarrow\downarrow$ pair (see main text). This Hamiltonian yields the following linearized self-consistency equation:
\be
\Delta_{{\vec K}\bar {\vec K}}(\vec k,\omega) = -\sum_{\vec k'\omega'} \frac{\Gamma_{\omega+\omega',\vec k+\vec k'}
	\Delta_{{\vec K}\bar {\vec K}}(\vec k',\omega')}{\omega'^2 + \epsilon^2_{{\vec K}+\vec k'}}
\ee
where we have accounted for the frequency and momentum dependence of the pairing vertex $\Delta_{{\vec K}\bar {\vec K}}$ since $\Gamma$ has a strong frequency and momentum dependence. Below we look for the leading pairing channel through symmetry analysis. 

To this end, we proceed as follows: we first identify the symmetries in biased Bernal bilayer graphene. Based on that, we can write down all symmetry-allowed pairing channels $\Delta_{{\vec K}\bar {\vec K}}(\vec k)$. Then identify channels that maximally benefit from the singularity of $\Gamma_{{\vec K}\bar {\vec K}}$. Finally, we work out the BCS pairing self-consistency equation in these (degenerate) channels and discuss the higher-order effects which select one channel from them.

The biased bilayer graphene has the following symmetries: a rotation $C_3$ around an axis perpendicular to the graphene plane, a mirror reflection $\mathcal{M}_y$ (a mirror lying in the $yz$-plane) that maps ${\vec K}$ to $\bar {\vec K}$. In addition, there is the time-reversal symmetry in the absence of the applied $B$ field. In the presence of $B_\parallel$ which couples only to spin, there is an orbital time-reversal symmetry left, which is denoted by $\mathcal{T}$. We note parenthetically that the inversion symmetry and the twofold rotation that swaps AB sites are both broken by the transverse field $D$. 

To understand the symmetry of $\Delta_{{\vec K}\bar {\vec K}}$, we need to understand the symmetry of the Hamiltonian in one valley. The $C_3$ operator, which maps $\psi_{{\vec K}+\vec k} \rightarrow \psi_{{\vec K} + C_3 \vec k}$ is preserved in each valley whereas the mirror symmetry $\mathcal M_y$ and time reversal $\mathcal T$ are both individually broken in each valley. However, the Hamiltonian is invariant under a combination of these two symmetry operators $\mathcal{M}_y \mathcal{T}$. In sum, the symmetry group is given by:
\be
G = \left\{1, C_3, C_3^2, \mathcal{M}_y \mathcal{T}, C_3 \mathcal{M}_y \mathcal{T} C_3^{-1}, C_3^{-1} \mathcal{M}_y \mathcal{T} C_3 \right\}
\ee

Therefore, the gap function $\Delta_{{\vec K}\bar {\vec K}}(\vec k)$ that respects the symmetry should obtain an overall phase under $C_3$ and $\mathcal{M}_x \mathcal{T}$, so we can write:
\bea\label{eq:Delta_symmetry}
&&  \Delta_{{\vec K}\bar {\vec K}}(C_3k) =  e^{i\phi_1} \Delta_{{\vec K}\bar {\vec K}}(\vec k), \quad \phi_1 = 0,\pm \frac{2\pi}{3},\\
&&  \Delta_{{\vec K}\bar {\vec K}}(\mathcal{M}_y \mathcal{T}\vec k) = e^{i\phi_2} \Delta_{{\vec K}\bar {\vec K}}(\vec k),\quad \phi_2 = \pm \pi.
\eea
We find that there are six possibilities in total: $\phi_1$ can take three values, whereas $\phi_2$ can take two values.

\section{The leading pairing channel}\label{sec:leading channel}

In this section, we argue that $p$-wave spin-triplet channel is the leading one in BBG based on symmetry classification. A similar result is obtained in Ref.\cite{chatterjee2022inter} through numerical calculation.

Which pairing channel is the strongest? To answer this question, we need to find the channels that maximally benefit from the singularity of pairing interaction $\Gamma_{\vec k \vec k'}$. Given that the singularity of $\Gamma_{\vec k \vec k'}$ occurs at $\vec k=-\vec k$, and that the SC gap function $\Delta_{{\vec K}\bar {\vec K}}(\vec k)$ is nonzero only for $\vec k$ near the Fermi surface, the strongest pairing channel is dominated by $\Delta_{{\vec K}\bar {\vec K}}$ at the momentum $\vec Q$ such that $\vec Q$ and $-\vec Q$ are both on the Fermi surface. To satisfy this, one way is to find a $\vec Q$ on the Fermi surface so that it is related to $-\vec Q$ through point group symmetry operation. Clearly, $C_3$ cannot relate $\vec Q$ and $-\vec Q$, but $\mathcal{M}_y \mathcal{T}$, $C_3 \mathcal{M}_y \mathcal{T} C_3^{-1}$ and $C_3^{-1} \mathcal{M}_y \mathcal{T} C_3$ can achieve this goal. As a result, we find six points $\pm Q_i$ ($i=1,2,3$) that satisfy this requirement, in which three $\vec Q_{i}$'s are related by $C_3$ rotation.  Alternatively, we can determine such $\vec Q$'s geometrically by Fig.\ref{fig:FS and valley-crossing points}c) in the main text, where $\vec Q$'s are given by the six intersections of superimposed Fermi surfaces in valley ${\vec K}$ and $\bar {\vec K}$.

It is convenient to describe the structure of the pairing function $\Delta_{{\vec K}\bar {\vec K}}$ using a valley-crossing-point model, where we define a gap function in each valley-crossing points $\pm \vec Q_i$ as $\Delta_{{\vec K}\bar {\vec K};\pm \vec Q_{i}}(\vec k)$. The electron energy near valley-crossing point $\vec Q_i$ is modeled as:
\be
\epsilon_{\pm i,\vec k} = v_F \vec{n}_{\pm i} \cdot 
\vec k+ \frac{|\vec{n}_{\pm i} \times \vec k	|^2}{2m_\perp} 
\ee
where $\vec n_i$ is the unit normal vector of Fermi surface at $Q_{i}$. In this model, we define a gap function in each valley-crossing point $\pm \vec Q_i$ as $\Delta_{{\vec K}\bar {\vec K};\pm Q_{i}}(\vec k)$. According to the symmetry constraint given in Eq.\eqref{eq:Delta_symmetry}, they satisfy
\be\label{eq:Delta_symmetry_valley-crossing points}
\Delta_{{\vec K}\bar {\vec K};\vec Q_{i}}(\vec k) = e^{i\phi(g)} \Delta_{{\vec K}\bar {\vec K};g \vec Q_{i}}(gk)), \quad g \in G
\ee
where 
\be
\phi(C_3) = 1, e^{\pm i\frac{2\pi}{3}}, \quad \phi(\mathcal{M}_y \mathcal{T}) = \pm 1, 
\ee
Using this model, we study the pairing problem. The self-consistency relation of pairing in this model can be written as
\be\label{eq:self-consistency_valley-crossing points_ij}
\Delta_{{\vec K}\bar {\vec K};\vec Q_{i}}(\vec k,\omega) = -\sum_{\vec k'\omega'j,\pm} \frac{\Gamma_{\omega+\omega',\vec Q_{i}+\vec k \pm \vec Q_{j}+\vec k'}
	\Delta_{{\vec K}\bar {\vec K};\pm \vec Q_{j}}(\vec k',\omega')}{\omega'^2 + \epsilon^2_{\pm j,\vec k'}}, 
\ee
Here $i,j = 1,2,3$. As a reminder, according to Eq.\eqref{eq:Gamma}, $\Gamma_{\omega,\vec k}$ diverges at $\omega=0,\vec k=0$. Therefore, below we neglect the $j\neq i$ term, deferring the discussion of these subleading interactions to later. Then Eq.\eqref{eq:self-consistency_valley-crossing points_ij} is simplified as follows:
\be\label{eq:self-consistency_valley-crossing points_ii}
\Delta_{{\vec K}\bar {\vec K};\vec Q_{i}}(\vec k,\omega) = -\sum_{\vec k'\omega'} \frac{\Gamma_{\omega+\omega',\vec k+\vec k'}\Delta_{{\vec K}\bar {\vec K};-\vec Q_{i}}(\vec k',\omega')}{\omega'^2 + \epsilon^2_{-i,\vec k'}},
\ee
Since the different valley-crossing points with $i=1$, $2$, and $3$ are decoupled under this approximation, in the analysis below, we only need to focus on one pair of valley-crossing points $\pm \vec Q_i$. For conciseness of the notation, below we refer to these two valley-crossing points as $\pm \vec Q$, and suppress the index $i$. 

Next, since the interaction $\Gamma_{\omega,\vec k}$ is positive-valued, pairing can only be generated in channels where the SC gaps in a pair of valley-crossing points related by $\mathcal{M}_y$ (i.e. in valley-crossing points $\vec Q_i$ and $-\vec Q_i$) are of opposite signs:
\be\label{eq:Delta_odd}
\Delta_{{\vec K}\bar {\vec K};\vec Q}(0,\omega) \Delta_{{\vec K}\bar {\vec K};-\vec Q}(0,-\omega) < 0
\ee
where the minus sign converts a repulsion to a pure attraction. 

In the analysis below, we only focus on the even-in-frequency pairing channels, i.e. the channels satisfying
\be \label{eq:Delta_even-in-frequency}
\Delta_{{\vec K}\bar {\vec K};\vec Q}(\vec k,\omega) = \Delta_{{\vec K}\bar {\vec K};\vec Q}(\vec k,-\omega).
\ee
This is justified because the odd-in-frequency channels are usually weaker since the gap functions in these channels are constrained by the odd-frequency requirement $\Delta(\vec k,0)=0$.

Comparing with the symmetries of all possible channels described in Eq.\eqref{eq:Delta_symmetry_valley-crossing points}, we find that the requirements Eq.\eqref{eq:Delta_odd} and Eq.\eqref{eq:Delta_even-in-frequency} can be simultaneously achieved in channels with $\phi(\mathcal{M}_y\mathcal{T})=-1$ (see Eq.\eqref{eq:Delta_symmetry_valley-crossing points}), i.e.
\be\label{eq:p wave}
\Delta_{{\vec K}\bar {\vec K};\vec Q}(\vec k,\omega) = - \Delta_{{\vec K}\bar {\vec K};-\vec Q}(\mathcal{M}_y \mathcal{T}\vec k,\omega).
\ee
We call such channels ``$p$-wave" channels. 

\section{Solving $\Delta_{{\vec K}\bar {\vec K}}(\vec k,\omega)$ and $T_{\rm c}$ in valley-crossing points model}\label{sec:Tc}
In this section, we analyze the self-consistency equation Eq.\eqref{eq:self-consistency_valley-crossing points_ii} and solve the momentum and frequency dependence of gap function $\Delta_{{\vec K}\bar {\vec K};\vec Q}(\vec k,\omega) $ for $p$-wave pairing channel. In literature such as Ref.\cite{chatterjee2022inter} this problem is analyzed numerically. Here we present an analytical solution that is obtained under approximations.

Plugging Eq.\eqref{eq:Gamma} and Eq.\eqref{eq:p wave} into Eq.\eqref{eq:self-consistency_valley-crossing points_ii} we get:
\be\label{eq:self-consistency_valley-crossing points_2}
\Delta_{{\vec K}\bar {\vec K};\vec Q}(\vec k,\omega) = \sum_{\vec k'\omega'} \frac{\Delta_{{\vec K}\bar {\vec K};\vec Q}(\vec k',\omega')}{\omega'^2 + \epsilon^2_{-i,\vec k'}} \frac{U}{ \kappa|\omega+\omega'|+ l_0^2 (\mathcal{M}_y \mathcal{T}\vec k+\vec k')^2+\delta^2}, 
\ee
We identify two momentum scales $k_\omega$ and $k_\delta$, where $k_\omega$ is set by the $\frac{1}{\omega^2+\epsilon^2_{\vec k'}}$ factor, and $k_\delta$ is given by the interaction:
\be
k_\omega = \frac{\omega}{v_F}, \quad k_\delta = \frac{\delta}{l_0}. 
\ee
Let $k_\perp$ be the momentum component perpendicular to Fermi velocity at ${\vec K}+\vec Q$, and $k_\parallel$ be the momentum component along Fermi velocity at ${\vec K}+\vec Q$, then $\Delta_{\vec K \bar{\vec K}}(\vec k)$ is nonzero only for $\vec k$ inside the following range:
\be\label{eq:chracteristic momenta}
|k_\perp|\lesssim k_\delta, \quad |k_\parallel|\lesssim \min \lp k_\omega, k_\delta \rp
\ee
This defines a characteristic frequency $\omega_0 = \delta v_F/l_0$. For $\omega>\omega_0$, the summation of $k_\parallel$ in Eq.\eqref{eq:self-consistency_valley-crossing points_2} is cut off at $k_\delta$. For $\omega<\omega_0$, this summation is cut off at $k_\omega$. 

To solve the self-consistency equation Eq.\eqref{eq:self-consistency_valley-crossing points_2} analytically, below we first show that, in this equation, the contribution from the regime of $\omega>\omega_0$ can be safely neglected. To see this, we analyze the contribution to the right-hand side (RHS) of Eq.\eqref{eq:self-consistency_valley-crossing points_2}. In that, the integral over $k_\parallel$ is cut off at $k_\delta$ (see Eq.\eqref{eq:chracteristic momenta}). Therefore, for an $\omega'$ in this regime, one finds $\omega'\gg \epsilon_{-i,\vec k'}$. As a result, the high-frequency contribution to RHS of Eq.\eqref{eq:self-consistency_valley-crossing points_2} can be written as
\be\nonumber
{\rm RHS} = 
\sum_{\vec k'_\perp, \omega'>\omega_0} \frac{U\Delta_{{\vec K}\bar {\vec K};\vec Q}(\vec k',\omega')}{ \omega'^2\lb \kappa|\omega+\omega'| + l_0^2 (\mathcal{M}_y \mathcal{T}\vec k+\vec k')^2+\delta^2 \rb } 
\ee
This integral over $\omega$ will not give a log divergence, therefore, the $\Delta(\omega)$ above $\omega_0$ only generates an $O(\frac{\Delta}{\omega_0})$ correction to the value of $\Delta$ obtained in low-frequency regime $\omega<\omega_0$. So long as we focus on the regime of $\Delta\ll \omega_0$, this contribution can be safely neglected.

Therefore, below we only need to solve the self-consistency equation \eqref{eq:self-consistency_valley-crossing points_2} in the low-frequency regime $\omega<\omega_0$. 
In this regime, we have $k_\omega<k_\delta$.
Therefore, we can integrate out $\vec k'_\parallel$ in the factor $\frac{1}{\omega'^2 + \epsilon_{-i,\vec k'}^2}$, and get the following linearized gap equation:
\be\label{eq:gap equation low frequency}
\Delta_{{\vec K}\bar {\vec K};\vec Q}(\vec k,\omega) = 
\sum_{k'_\perp, \omega'<\omega_0} \frac{\Delta_{{\vec K}\bar {\vec K};\vec Q}(\vec k',\omega')}{2v_F |\omega'| } \frac{U}{\kappa|\omega+\omega'|+ l_0^2 (k_\perp^2+ { k'}_\perp^{2}-2k_\perp k_\perp'\cos \theta)+\delta^2}, 
\ee

Below we solve the self-consistency equation Eq.\eqref{eq:gap equation low frequency}. For simplicity, we neglect the $\kappa |\omega+\omega'|$ term in the denominator. This approximation will be justified at the end of the discussion. To proceed analytically, we replace the rest of the long denominator with the following separable form:
\be
\sqrt{ \lp 2(1-\cos\theta) l_0^2 {k}_\perp^2+\delta^2\rp \lp 2(1-\cos\theta) l_0^2 {k'}_\perp^2+\delta^2\rp}  
\ee
This substitution enables separating the momentum and frequency dependence in the gap function $\Delta_{{\vec K}\bar {\vec K}}(\vec k,\omega)$ as following
\be
\Delta_{{\vec K}\bar {\vec K};\vec Q}(\vec k,\omega) = \frac{\phi(\omega)}{ \sqrt{ 4 l_0^2 {k'}_\perp^2\sin^2\frac{\theta}{2}+\delta^2}},\label{eq:Delta_result}
\ee
Here $\phi(\omega)$ is the frequency-dependent part of $\Delta_{{\vec K}\bar {\vec K};Q}(\vec k,\omega)$, which is governed by the following self-consistency relation:
\be
\phi(\omega) = 
\sum_{\omega'<\omega_0} \frac{\phi(\omega')}{2v_F |\omega'| } \sum_{k'_\perp}\frac{U}{ 4 l_0^2 {k'}_\perp^2\sin^2\frac{\theta}{2}+\delta^2}. 
\ee
Carrying out the summation over ${k'}_\perp^{2}$ on right hand side, we find
\be
\phi(\omega) = 
\sum_{\omega'<\omega_0} \frac{\phi(\omega')}{2v_F |\omega'| } \frac{U}{4l_0\delta\sin\frac{\theta}{2}}. 
\ee
This gives a critical temperature of
\be\label{eq:Tc}
T_{\rm c} = 2\omega_0 e^{-\frac{1}{\lambda}}, \quad \lambda = \frac{U}{8 v_Fl_0\delta\sin\frac{\theta}{2}}.
\ee
As a reminder $\omega_0 = \delta v_F / l_0$. We find a maximal SC critical temperature of $T_{\rm c,max}= \frac{U}{2l_0^2}$, which is achieved when $\delta = \frac{U}{4v_Fl_0}$. Using the Stoner criterion $U \nu_0=1$ where $\nu_0$ is the density of states, we find the maximal $T_{\rm c}$ is comparable to Fermi energy, which is much larger than the $T_{\rm c}$ predicted by the valley-polarization (VP) fluctuations \cite{dong2022spin}. Therefore, it is reasonable to consider this channel as the leading pairing channel. We note that the $T_{\rm c}$ predicted here is obtained when only the effects of IVC fluctuations are accounted for.  
%\addQ{\sout{We will discuss the effects that suppress the Tc, such as valley polarization fluctuations.}}

Below we check the validity of our analysis. As a reminder, in the analysis above we ignored the $\kappa|\omega+\omega'|$ in Eq.\eqref{eq:gap equation low frequency}. Is this approximation valid? It is valid when the frequency in Eq.\eqref{eq:gap equation low frequency} is below a threshold $\omega_*$, which is defined as $\omega_* = \delta^2/\kappa \sim 4v_F^2 \delta^2/U$. However,  in Eq.\eqref{eq:gap equation low frequency} the frequency $\omega'$ is summed up to $\omega_0$. Therefore, this approximation is acceptable only when $\omega_0\lesssim \omega_*$, for that we need $\delta \gtrsim U/4v_Fl_0$, which happens to be the optimal value of $\delta$ where $T_{\rm c}$ is maximized. 
For $\delta < U/4v_Fl_0$, we expect the $T_{\rm c}$ to be further suppressed as compared to the prediction of Eq.\eqref{eq:Tc}. We can estimate the $T_{\rm c}$ in this small $\delta$-regime by replacing the bandwidth $\omega_0$ in Eq.\eqref{eq:Tc} with a lower threshold frequency $\omega_*$.

\end{widetext}
	
\end{appendix}

\end{document}